\documentstyle[twoside,epsf,fleqn,espcrc2]{article}

\newcommand{\AmS}{{\protect\the\textfont2
  A\kern-.1667em\lower.5ex\hbox{M}\kern-.125emS}}

\title{The Tokyo Axion Helioscope Experiment\thanks{Talk given at the
Fifth IFT Workshop: Axions, 13-15 March 1998, Gainesville, Florida}}
\author{M.~MINOWA\address{Department of Physics,
School of Science, University of Tokyo, 7-3-1 Hongo, Bunkyo-ku, Tokyo
113-0033, Japan}\address{RESCEU, Research Center for the Early
Universe, School of Science, University of Tokyo, 7-3-1 Hongo,
Bunkyo-ku, Tokyo 113-0033, Japan}, S.~Moriyama$^{\rm ab}$,
Y.~Inoue$^{\rm b}$\address{ICEPP, International Center for Elementary
Particle
Physics, University of Tokyo, 7-3-1 Hongo, Bunkyo-ku, Tokyo 113-0033,
Japan}, T.~Namba$^{\rm a}$, Y.~Takasu$^{\rm a}$, and
A.~Yamamoto\address{High Energy Accelerator Research
Organization(KEK), 1-1 Oho, Tsukuba, Ibaraki 305-0801, Japan}}

\begin{document}

\begin{abstract}
A preliminary result of the solar axion search experiment at the
University of Tokyo is presented.  We searched for axions which could
be produced in the solar core by exploiting the axion helioscope.  The
helioscope consists of a superconducting magnet with field strength
of 4\,Tesla over 2.3 meters.  From the absence of the axion signal we
set a 95\,\% confidence level upper limit on the axion coupling to two
photons $g_{a\gamma\gamma} < 6.0 \times 10^{-10} {\rm GeV}^{-1}$ for
the axion mass $m_a < 0.03$\,eV.  This is the first solar axion search
experiment whose sensitivity to $g_{a\gamma\gamma}$ exceeds the limit
inferred from the solar age consideration.
\end{abstract}

\maketitle

\section{INTRODUCTION}
The axion is a light pseudoscalar particle introduced to solve the
strong CP problem\cite{PQ77,WW,Physrep,DFSZ,KSVZ}.  However, the
theory of axion does not predict its mass.  The axion would be
produced in the solar core through the Primakoff effect if its mass is
a few electronvolts.

\begin{figure}[ht]
\begin{center}
\epsfxsize=7.5cm
\epsffile{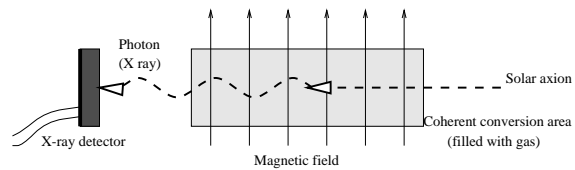}
\caption{Principle of solar axion detection}
\label{fig:principle}
\end{center}
\end{figure}

Sikivie\cite{Sikivie83} proposed an experiment to detect the axions
emitted by the sun using a system of a strong magnetic field and an
x-ray detector, called the axion helioscope.  In the magnetic field,
solar axions convert back to x-rays of black body radiation spectrum
with an average energy of about 4 keV by the inverse process.  The
conversion can be coherently enhanced by filling the conversion region
with dense gas\cite{Bibber89}.  
The principle of detection is illustrated in Fig.\ref{fig:principle}.

We constructed a dedicated superconducting magnet to search for the
solar axions.  We are going to adopt cold helium gas as conversion
medium with a temperature just above 4K, the boiling point at one
atmosphere.  With electron density of this medium, conversion of
axions with mass of 2.65\ eV gets enhanced coherently. However, we
tried without any conversion medium at first because we need some time
to develop the gas container.  We report on the results of this first
helioscope run without conversion gas.
 
\section{AXION HELIOSCOPE}
\begin{figure}[ht]
\begin{center}
\epsfxsize=7.5cm
\epsffile{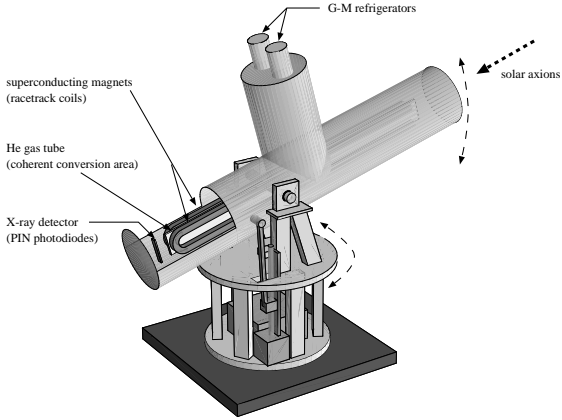}
\caption{Schematic view of the axion helioscope}
\label{fig:helioscope}
\end{center}
\end{figure}

The axion helioscope consists of a superconducting magnet\cite{Sato},
x-ray detectors, and an altazimuth to direct the helioscope toward the
sun.  A schematic illustration
of the axion helioscope is
shown in Fig.\ref{fig:helioscope}.

The superconducting magnet has a field strength of 4\ Tesla over
effective length of 2300 mm. With a help of two Gifford-McMahon
refrigerators attached on its top, it is cooled down to 5.2\ K without
any liquid helium. The magnet consists of two split racetrack
superconducting coils, which are connected to the refrigerators by
conduction rods.  These coils are installed in a cylindrical vacuum
chamber.

It takes 14 days to achieve the temperature needed to operate
the magnet if we start from the room temperature.  
Excitation of the magnet takes one hour to reach the field strength of 
4\ Tesla. The field is then maintained by a persistent current of 268
amperes.  
Main parameters of the magnet are summarized in the Table\ref{tab:paramofmag}.

\begin{table*}[ht]
  \begin{center}
    \leavevmode
    \begin{tabular}{ll}
      \hline
      Coil & \\
      \hline
      \ \ Shape & racetrack\\
      \ \ Quantity & 2\\
      \ \ Winding thickness & 64\,mm\\
      \ \ Winding width & 50\,mm\\
      \ \ Winding cross section area & $3.2\times 10^3\,\rm mm^3$\\
      \ \ Inner/outer radius of both ends of coil & 50/100\,mm\\
      \ \ Liner length & 2100\,mm\\
      \ \ distance between coils & 20\,mm\\
      \ \ volume & 14.95$\times10^4$\,mm$^3$/coil\\
      \ \ weight & 133.5\,kg/coil\\
      \hline
      Winding coil & \\
      \hline
      \ \ layer & 54/coil\\
      \ \ turn & 33.5/coil\\
      \ \ number of turns & 1809/coil\\
      \hline
      Operation current & 268\,A (336\,A)\\
      \hline
      Central magnetic field & 4.0\,T (5.0\,T)\\
      \hline
      Max magnetic field & 5.74\,T (7.18\,T)\\
      \hline
      Inductance & 15.5\,H\\
      \hline
      Stored energy & 560\,kJ (875\,kJ)\\
      \hline
      Weight (4\,K part) & 670\,kg\\
      \hline
      GM-refrigerator & \\
      \hline
      \ \ Quantity & 2\\
      \ \ capacity & 1st: 20\,W @ 40\,K\\
      & 2nd: 0.5\,W @ 4.2\,K\\
      \hline
    \end{tabular}
  \end{center}
  \caption{Main parameters of the magnet.
    Designed parameters are shown in parentheses.}
  \label{tab:paramofmag}
\end{table*}

The magnet is supported on the altazimuth, with which it can be turned
by 360 degrees around the vertical axis, and $\pm$28 degrees above and
below the horizontal plane. Averaging over one year in Tokyo, the sun
is within $\pm$28 degrees of altitude for about 50\% of the time. When
the sun is out of this region, we take background data. The mounting
is controlled by a computer to track the sun. The horizontal and
elevational origin was determined by a precision theodolite.  We
determined the north direction by observing the star called
$\beta$-Ori.  The horizontal level was determined by a spirit level.

The errors of the tracking were found to be smaller than $\pm
0.50$\,mrad in the azimuthal direction and $\pm 0.45$\,mrad in the
elevational direction.  Since the aperture of the helioscope is $\pm
5$\,mrad, these errors are small enough. 

For the axion-converted x-ray detection, we use 16 pieces of PIN
photodiodes, Hamamatsu S3590-06.  9 pieces out of 16 were actually
used for the first measurement because of unexpected damages during 
the cooling. These are
windowless photodiodes with active area of $\rm9\,mm\times9\,mm$ each
and thickness of 500 $\mu$m.  The typical capacitance of the diode is
30\ pF with a reverse bias voltage of 100\ V.  It is originally supplied
on a ceramic substrate. However, we ordered special diodes with
substrates of Kapton and Teflon because we observed that the ceramic
had detectable radioactive contaminations in it.
The temperature of the diodes is kept around 50K by the first stage 
of the refrigerators.

Energy calibration was done by irradiating x or $\gamma$ rays of
checking sources before the detector installation.
The energy resolution of the PIN photodiodes was 0.34\,keV in the
energy region of the present interest as determined by test
pulses.
The detection efficiency was estimated with 5.9\,keV
x-rays from $^{55}$Fe of known activity.  
It was found that the absolute detection efficiency was at least 0.5
at 4\,keV, and more than 0.9 at 10\,keV.

\section{MEASUREMENT and ANALYSIS}
The first measurement was done from 26th till 31st December 1997.  We
tracked the sun with the axion helioscope so far as its altitude is
within 28 degrees above and below the horizon.  The earth can be
practically considered to be transparent to the axion.  A total time of
$1.9\times 10^5$ sec was dedicated to the tracking run.  We measured
the background data during the rest of the time, which amounted to
$2.0\times 10^5$ sec.

The output signals from the preamplifiers are fed into shaping
amplifiers for the trigger.  On the other hand, the waveform of the
preamplifier outputs are digitized over 50\,$\mu$sec before and after
a trigger and recorded with a sampling period of 0.1\,$\mu$sec.  The
stored waveforms are then shaped digitally in a computer.

\begin{figure}[ht]
\begin{center}
\epsfxsize=7.5cm
\epsffile{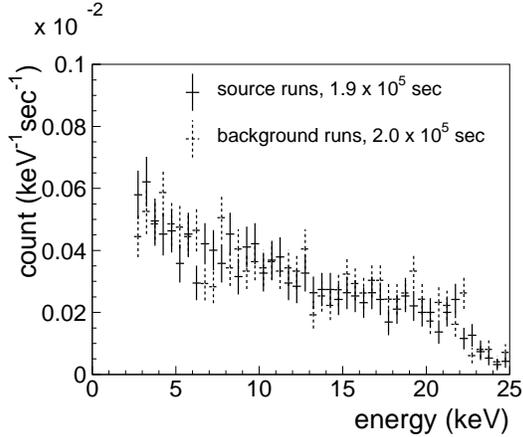}
\caption{Energy spectra of source runs and background runs}
\label{fig:spectra}
\end{center}
\end{figure}

After event selection cuts to eliminate cosmic muon events, we got
energy spectra of source runs and background runs, each of which is the
sum of the data from nine PIN photodiodes.  They are shown in
Fig.\ref{fig:spectra}.

Since the axion signal can be observed only in the source spectrum, we
subtracted the background contribution from it.  We then fit the
theoretical spectrum to it.  The fitting region is
restricted to the energy region between 4 and 14\,keV, where most
events are concentrated and the
trigger efficiency is almost 100\%.

The conversion probability, $p_{a\rightarrow\gamma}$, is written by a
Fourier transformation of the magnetic field\cite{Bibber89},

\begin{equation}
p_{a\rightarrow\gamma}=\left|\int_0^L\frac{g_{a\gamma\gamma}}{2}
B(x,y,z)\exp(izq){\rm d}z \right|^2,
\end{equation}

where $g_{a\gamma\gamma}$ is the axion coupling constant to two
photons, $B$ and $L$ are the magnetic field and its length,
$q=m_a^2/2E$ is the momentum transfer, $z$ is a coordinate along the
magnet axis, and $x$ and $y$ are coordinates perpendicular to the
$z$ axis.  In the fitting, the coupling constant $g_{a\gamma\gamma}$
is left free.  Thus obtained $p_{a\rightarrow\gamma}$ is multiplied by
the axion differential flux, the detection efficiency and the trigger
efficiency.  We then get the theoretical spectrum after convolution
with the detector response function with the energy resolution.

\begin{figure}[ht]
\begin{center}
\epsfxsize=7.5cm
\epsffile{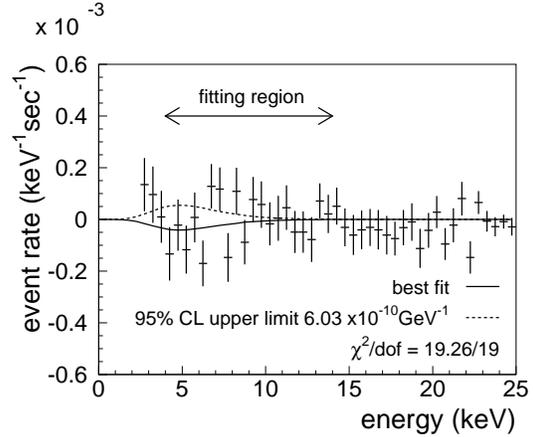}
\caption{Fitting of the axion spectrum with experimental data with
$m_a$=0.001\,eV.}
\label{fig:fit1}
\end{center}
\end{figure}

\begin{figure}[ht]
\begin{center}
\epsfxsize=7.5cm
\epsffile{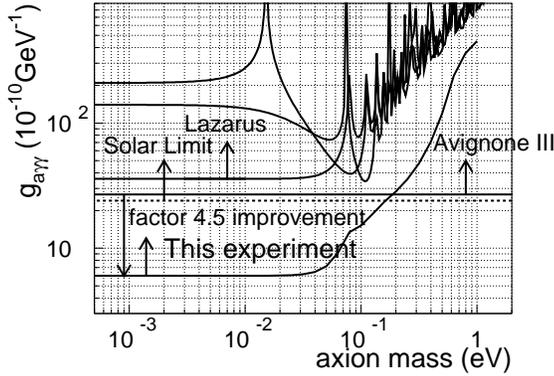}
\caption{The 95\% CL upper limits on $g_{a\gamma\gamma}$.  
Also shown are the solar limit, the limits by Avignone 
{\it et al.}\cite{Avignone} and Lazarus {\it et al.}\cite{Lazarus}.}
\label{fig:results}
\end{center}
\end{figure}

The fitting was done assuming a certain value for $m_a$.
Fig.\ref{fig:fit1} shows an example for $m_a=0.001$\,eV.  The solid
line corresponds to the best fit and the dashed line the 95\%
confidence level upper limit.  Then the fitting is repeated with
various values for $m_a$.  We thus obtain 95\% CL upper limits as a
function of $m_a$ in the range from 0.001 to 1\,eV as shown in
Fig.\ref{fig:results}.  In the same figure, upper limits from earlier
experiments by other groups as well as the limit inferred from the
solar age consideration are also shown.  This is the first solar axion
search experiment which has sufficient sensitivity to
$g_{a\gamma\gamma}$ below the solar limit. The present experiment
gives factor of 4.5 more stringent limit than the recent limit by
Avignone {\it et al.}\cite{Avignone} in the region $m_a<$0.03\,eV.

\section{CONCLUSION}
We searched for axions which could be produced in the solar core using
the axion helioscope, which consists of a super conducting magnet with
field strength of 4\,Tesla over 2.3 meters.  From the absence of the
axion signal we set a 95\,\% confidence level upper limit on the axion
coupling to two photons $g_{a\gamma\gamma} < 6.0 \times 10^{-10} {\rm
GeV}^{-1}$ for the axion mass $m_a < 0.03$\,eV.  This is the first
solar axion search experiment whose sensitivity to $g_{a\gamma\gamma}$
exceeds the limit inferred from the solar age consideration.  The
present limit gives a factor of 4.5 improvement over the formerly best
limit in the region $m_a<$0.03\,eV.

\section{PROSPECTS}
After completion of the development of the gas container, in which the
cold conversion gas is to be filled, we will start the next
measurement.  In this new measurement we should have sensitivity in
$m_a$ range between 0.03 and 2.6\,eV, especially we can reach the
sensitivity predicted by the hadronic axion models\cite{KSVZ} around
$m_a=2.6$\,eV.

\section*{ACKNOWLEDGEMENTS}
We are indebted to S.~Mizumaki who fabricated the persistent current
switch of the magnet.  We are also thankful to Y.~Sato for her support 
in operating the superconducting magnet in the beginning stage of the
experiment.
We further thank S.~Otsuka for his support of our engineering work.
We express gratitude to F.~Shimokoshi who designed the original form
of the preamplifier circuit.  We express gratitude to Y.~Makida and
K.~Tanaka who helped us to prepare various instruments associated with 
the magnet.
We are indebted to H.~Hara for his contribution in the first stage of
the development of the x-ray detectors.

This research is supported by the Grant-in-Aid for COE research by the
Japanese Ministry of Education, Science, Sports and Culture, and also
by the Matsuo Foundation.

\end{document}